# CONSIDRERATION OF HTS RAPID-CYCLING MAGNET FOR STAGED MUON ACCELERATION

Henryk Piekarz [†]
Fermi National Accelerator Laboratory, Batavia, Illinois 60510,

*Abstract*

A possible application of the HTS (*Re*BCO) rapid-cycling dipole magnet [1] in the first stages of muon acceleration consisting of Recirculating Linear Accelerator (RLA) and followed-up first Rapid Cycling Synchrotron (RCS-1) is presented. The projected *Re*BCO magnet hysteresis loss is discussed in terms of the liquid helium coolant choice and the option with the striated *Re*BCO conductor for the further significant reduction of the hysteresis power loss.

## INTRODUCTION

The short muon lifetime requires rapid acceleration to minimize the decay losses on the way to the muon collider injection energy. As the muon lifetime is the shortest at low energies the muon acceleration is arranged in stages [2] consisting of the Recirculating Linear Accelerator (RLA) for the fastest initial acceleration followed with the sequence of multiple Rapid Cycling Synchrotrons (RCSs) to reach the beam injection energy of the Muon Collider. The RLA consists of the linear accelerator (LA) where the muons gain energy, and the Returning Synchrotrons (RSs) at both ends of the LA to facilitate subsequent multiple accelerations.

The RCSs accelerator magnetic field determines the ring length for the muon exit energy, and the field ramping speed sets the acceleration period which should be as close as possible to the muon total circulation time. For this reason, the RCS's must use the dipole magnets of a very high ramping speed. The RSs dipole magnet can operate in the DC mode, but the AC mode allows for a very significant power saving, especially for the Muon Collider proposed operation only at 5 Hz and the muon circulation times in the range of small fraction of a millisecond. Consequently, the RS synchrotrons should also be based on the rapid-cycling magnets.

## ARRANGEMENT OF RLA and RCS-1

The proposed arrangement of the RLA is shown in Fig. 1. The RLA comprises of the dual polarity 0.4 GeV muon source, the 8 GeV LA and four RS synchrotrons at each end of LA. The 0.4 GeV $\mu^+$ and $\mu^-$ muons enter the 8 GeV LA in its centre and while traveling in opposite directions are accelerated to 4 GeV. Both muon species with help of the RSs 1-2-3-4-5-6-7 pass 7 times through the LA to be accelerated to the total of 60 GeV. Each muon specie uses its own RS1 and RS7 while sharing the RSs 2-3-4-5-6. Based on the 8 GeV Linac [3] currently under construction at Fermilab (SRF 1300 MHz, 33.7 MV/m, 20 cryomodules @ 12.5 m), the 8 GeV LA beam path is about 260 m. With the RSs 2 T top B-field the approximated muon paths and the flight times are given in Table 1. The estimated muon path in the RLA is ~ 6000 m and the flight time is ~ 20 µs.

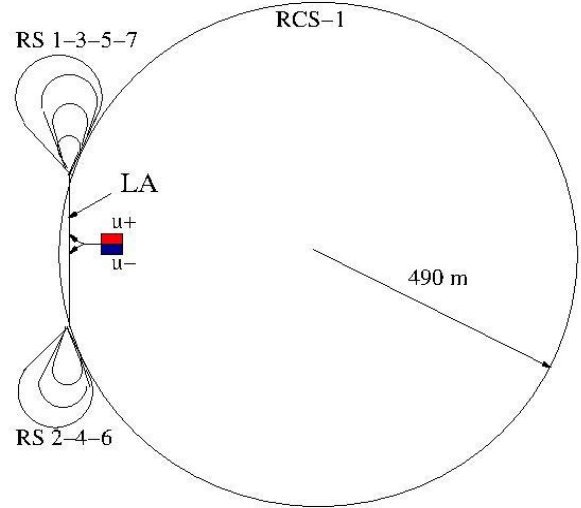

Figure 1: Arrangement of RLA and RCS-1 with 8 GeV LA. For clarity only RS 1 and RS 7 paths for $\mu^+$ muons are shown.

Table 1. Approximated muon paths and flight times

| RLA components | Beam path | Flight time |
|---|---|---|
| [GeV] | [m] | [µs] |
| RS1    4 | 192 | 0.64 |
| RS2   12 | 426 | 1.42 |
| RS3   20 | 509 | 1.70 |
| RS4   28 | 593 | 1.98 |
| RS5   36 | 677 | 2.26 |
| RS6   44 | 760 | 2.53 |
| RS7   52 | 844 | 2.81 |
| **RS 1-7**   4 - 52 | 4001 | 13.14 |
| **LA**   0.4 - 60 | 1875 | 5.50 |
| **RLA**  0.4 - 60 | 5876 | 18.64 |

The fast-ramping dipole magnet is constructed with the Fe3%Si laminations for which the B-field-to-current linear response is up to 1.7 T with the full saturation at 2 T limiting the range of muon exit energy for a given accelerator circumference. The energy range of the RCS accelerator can be expanded, however, by increasing the accelerator average magnetic field. This can be achieved by combining the rapid-cycling low-field ac magnets with the high-field dc magnets [4]. For the average accelerator B-field of 3 T the combination of the 1.7 T rapid-cycling magnets with the 8 T dc magnets the accelerator circumference of the 400 GeV beam is 3070 m assuming the magnet packing factor of 90%. For the relativistic muons the flight time per single circulation is then ~ 10 μs, and the number of circulations with the 8 GeV/turn is 42. This leads to the total muon flight tine of 420 μs to reach 400 GeV and to the required accelerator ramping magnetic field of ~ 8 kT/s.

## *Re*BCO DIPOLE MAGNET FOR RS & RCS SYNCHROTONS

As both the RS and RCS synchrotrons require operations at the time scale shorter than $10^{-3}$ s they can be based on the same type of the rapid-cycling magnet. The characteristic response of the B-field to the energizing current for the 2 T field in the 30 mm beam gap of the Fe3%Si dipole magnet is shown in Fig.2. The B-field saturation period can be used for the operation of the RS synchrotrons while the +/- 1.7 T ramp for the RCS synchrotrons. At the ramping rate of 8 kT/s the accelerator period for the +/- 1.7 T field ramp is 0.425 $10^{-3}$ s, and flat magnetic field period is $10^{-3}$ s (duty factor ~ 0.5 % @ 5 Hz). As the muon total circulation time through the chain of the RSs is ~ 20 $10^{-6}$ s they could operate at a much faster ramp rate. But as the cycle is determined by B-field saturation the RSs magnets will use the similar operational cycle as the RCSs. Consequently, the ReBCO rapid-cycling dipole magnet can be applied for both, the RS and RCS accelerators.

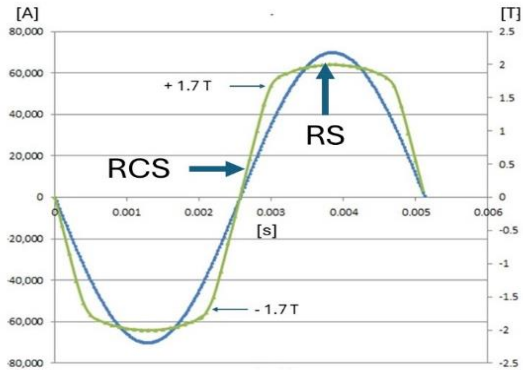

Fig. 2 Response of magnet B-field to the sinewave current for the Fe3%Si with 30 mm beam gap. The RCS operations are in the range of +/- 1.7 T while the RS uses the ~ 2 T B-field flat top.

The *Re*BCO conductor hysteresis is the main source of the magnet cable power loss. This power is proportional linearly [5] to the magnetic field crossing the cable space which makes it independent of the B-field ramping rate. This feature makes the *Re*BCO magnet suitable for the application for the muon acceleration.

From ref [1] the conceptual design of 2 T *Re*BCO dipole magnet with the 30 mm beam gap is shown in Fig. 3, and the *Re*BCO cable design in Fig. 4. The two parallel 12 kA power supplies energize two 3-turn *Re*BCO cables for 72-kA magnet current. The six 6 kA cables, for a total of 36 kA, are arranged vertically to facilitate turns around the centre of the magnet core. The optimized cable placement for the minimal transverse magnetic field through the cable space is shown in Fig. 5. The calculated momentary hysteresis loss in ¼ magnet energized with 36-kA step-function current is ~7.6 J/m, 61 J/m for the full magnet and full ramp.

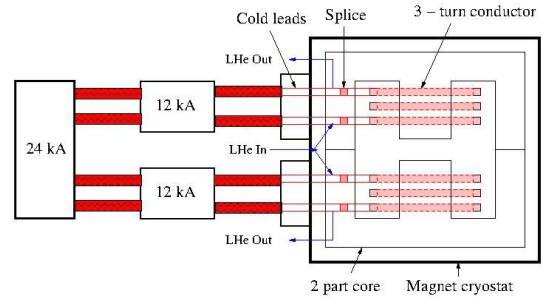

Fig. 3 Conceptual design of the *Re*BCO magnet with 2 coils of 3 conductor turn of 2 *Re*BCO cables in parallel around an H-core carrying 12 kA per turn, 6 kA per cable, in total 72 kA cable-turn per magnet.

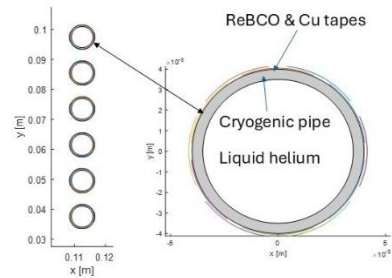

Fig. 4 *Re*BCO cable design for 6 kA. 12 tapes of 2 mm width wrapped around the cooling pipe of 8 mm OD in a single layer. Six cable sections are arranged in 3 turns of 2 cables in parallel per pole. 72 cable sections in total per magnet.

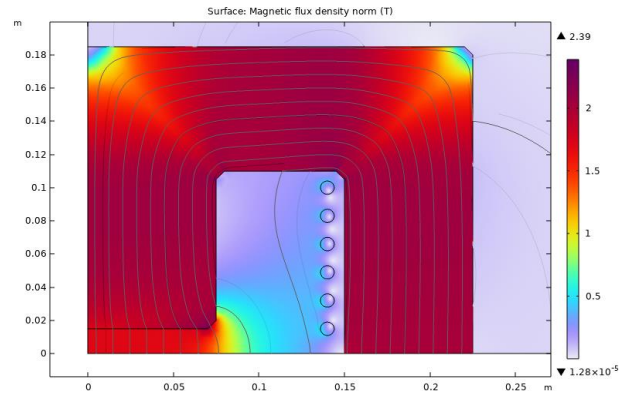

Fig. 5 B-field calculation for magnet core and 6 *Re*BCO cable sections. Averaged transverse magnetic field in cable space is 0.19 T.

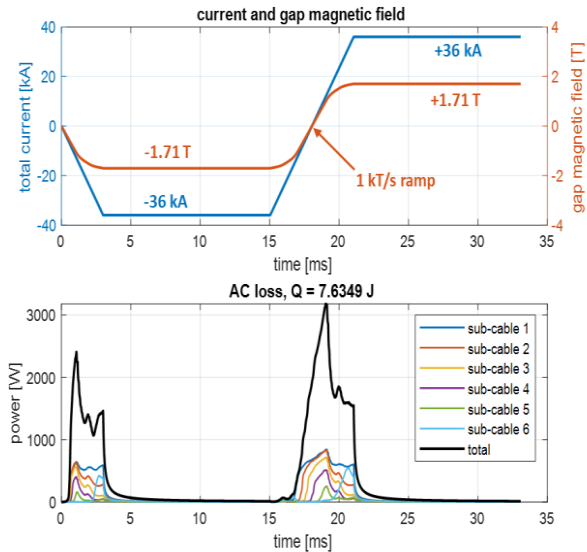

**Fig. 6.** Current in the six cables and gap magnetic field (top). Momentary hysteresis loss versus time in ¼ of the *Re*BCO magnet (bottom).

As outlined in detail in [1] this hysteresis loss can be much reduced with operations at 5.5 K. The helium high specific heat at 5.5 K reduces the temperature rise by about a factor of 6 relative to that at 15 K. And at the 5.5 K fewer *Re*BCO tapes are needed for the same operational current requiring in turn a smaller diameter of supporting cryogenic pipe and consequently minimizing cable exposure to the core descending magnetic field. In total, the hysteresis loss at 5.5 K can be reduced by a factor of 18 relative to that at 15 K, e.g. from 61 J/m down to about 3.4 J/m.

Further hysteresis loss reduction is possible by applying *Re*BCO tapes narrower than 2 mm, or by the striation of the *Re*BCO tape surface into the multiple filaments [6]. Several striation methods, such as laser ablation, mechanical scribing or chemical etching successfully created multiple filaments in the short samples of wide *Re*BCO tapes while preserving high retention of the critical current. As example, with the reported width of a typical groove in the (40-50) μm range, 4 grooves on the surface of 2 mm wide *Re*BCO tape will make 5 filaments reducing the hysteresis loss by a factor of 5 while preserving the critical current at ~ 90 % level.

The required cryogenic electric power depends on the Coefficient of Performance (CoP) and the cryogenic plant efficiency. The CoP for the operation at 5.5 K is 1.9% and 5% at 15 K. The cryogenic plant efficiency for a large-scale system (as needed for the muon accelerator) is about 20%. With these parameters the electric power for the operations with 2 mm *Re*BCO tapes is about 4.5 kW/m at 5.5 K and 30 kW/m at 15 K. The summary of the *Re*BCO cable hysteresis loss for 2 mm tapes without and with striation is presented in Fig. 7.

On notice, the 4.5 kW electric power for the magnet of 2 T, 10 kT/s at 5 Hz is below 8.6 kW of the FNAL Booster of 0.7 T, 30 T/s at 15 Hz [7].

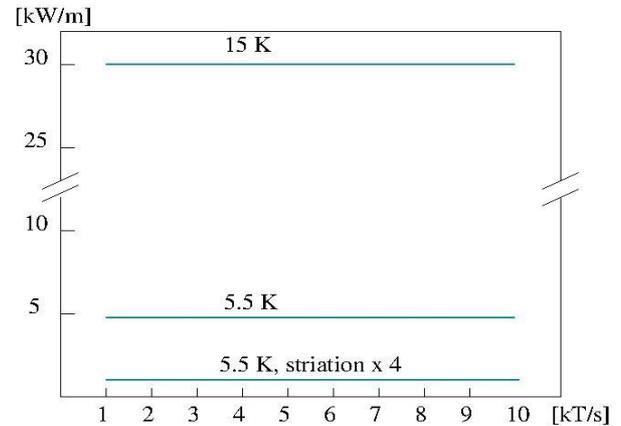

**Fig. 7.** 2T *Re*BCO magnet electrical cryogenic power for 2 mm tapes at 15 K and at 5.5 K without and with 4 50 μm grooves striation.

At high-ramping operation, the magnet conductor must withstand the repeatable high-voltage shock induced by the ramping high current. The inductance of the proposed *Re*BCO magnet is 48 μH/m which with the 12-kA energizing current generates voltage shock of 1700 V at 10 kT/s ramp rate, requiring proper cable electrical insulation. As outlined in [1], the *Re*BCO cable can be insulated from the magnet core with multiple ABS (Acrylonitrile Betadine Styrene) holders and the core itself insulated with the G10 spacers from the cryostat wall which is electrically connected to the ground. Such an insulation can withstand a voltage rise much higher than 2 kV.

## CONCLUSION

The *Re*BCO dipole magnet as designed in [1] is suitable for application in first stages of muon acceleration requiring the fastest possible ramp rates at low power loss.